\documentclass[prl,twocolumn,showpacs,superscriptaddress,a4paper]{revtex4}
\usepackage{amssymb}
\usepackage{epsfig}
\usepackage{amsmath}
\usepackage{graphicx}
\usepackage{bm}

\begin{document}

\title{Ground-state cooling of a micromechanical oscillator: comparing cold-damping and cavity-assisted cooling schemes}
\author{C. Genes}
\affiliation{CNISM and Dipartimento di Fisica, Universit\`{a} di Camerino, I-62032 Camerino (MC), Italy}
\author{D. Vitali}
\affiliation{CNISM and Dipartimento di Fisica, Universit\`{a} di Camerino, I-62032 Camerino (MC), Italy}
\author{P. Tombesi}
\affiliation{CNISM and Dipartimento di Fisica, Universit\`{a} di Camerino, I-62032 Camerino (MC), Italy}
\author{S. Gigan}
\affiliation{Institut f\"{u}r Experimentalphysik, Universit\"{a}t Wien, Boltzmanngasse 5,
1090 Wien, Austria and Institute for Quantum Optics and Quantum Information
(IQOQI), Austrian Academy of Sciences, Boltzmanngasse 3, 1090 Wien, Austria}
\author{M. Aspelmeyer}
\affiliation{Institut f\"{u}r Experimentalphysik, Universit\"{a}t Wien, Boltzmanngasse 5,
1090 Wien, Austria and Institute for Quantum Optics and Quantum Information
(IQOQI), Austrian Academy of Sciences, Boltzmanngasse 3, 1090 Wien, Austria}

\begin{abstract}
We provide a general framework to describe cooling of a micromechanical oscillator to its quantum ground state by means of radiation-pressure
coupling with a driven optical cavity. We apply it to two experimentally realized schemes, back-action cooling via a detuned cavity and
cold-damping quantum-feedback cooling, and we determine the ultimate quantum limits of both schemes for the full parameter range of a stable
cavity. While both allow to reach the oscillator's quantum ground state, we find that back-action cooling is more efficient in the good cavity
limit, i.e. when the cavity bandwidth is smaller than the mechanical frequency, while cold damping is more suitable for the bad cavity limit.
The results of previous treatments are recovered as limiting cases of specific parameter regimes.
\end{abstract}

\maketitle

\section{Introduction}
Cooling of mechanical resonators close to their quantum ground state has become an important topic for various fields of physics, such as
ultra-high precision measurements \cite{schwab}, the detection of gravitational waves \cite{GW}, and the study of the transition between
classical and quantum behavior of a mechanical system \cite{found}. It is also a prerequisite for any possible use of optomechanical systems for
quantum information processing \cite{prltelep,prl07}. Recently, various experiments have demonstrated significant cooling of the vibrational
mode of a mechanical resonator coupled to an optical cavity
\cite{cohadon99,karrai04,naik,gigan06,arcizet06,arcizet06b,bouwm,vahalacool,mavalvala,rugar,harris}. In these experiments cooling has been
achieved by exploiting in two different ways the radiation-pressure interaction between a mechanical mode and the intra-cavity field: i) by
back-action, or self-cooling \cite{brag}, in which the off-resonant operation of the cavity results in a retarded back action on the mechanical
system and hence in a ``self''-modification of its dynamics \cite{gigan06,arcizet06b,vahalacool,mavalvala,meystre}; ii) by cold-damping quantum
feedback \cite{Mancini98,courty,quiescence02}, where the oscillator position is measured through a phase-sensitive detection of the cavity
output and the resulting photocurrent is used for a real-time correction of the dynamics \cite{cohadon99,arcizet06,bouwm,rugar}.

We generalize and extend the previous treatments of these schemes~\cite{brag,Mancini98,courty,quiescence02,marquardt,wilsonrae} to the full
parameter range of a stable cavity by deriving the quantum steady state of the micromechanical oscillator in a linearized quantum Langevin
equation (QLE) approach. Comparing the two schemes we find that back-action cooling is more efficient in the good cavity limit, i.e. when the
cavity bandwidth is smaller than the mechanical frequency, while cold damping is more suitable in the opposite limit of a bad cavity. We also
show that, contrary to common belief, the feedback gain in cold-damping schemes is necessarily bounded by an upper limit to achieve quantum
ground-state cooling.

The paper is organized as follows. In Section II we describe the dynamics of the system in terms of linearized quantum Langevin equations. In
Sec. III and IV we evaluate the steady-state energy of the mechanical oscillator for the two cases of back-action cooling with a detuned cavity
and cold-damping feedback cooling. In Section V we conclude by comparing in detail the two cooling schemes.

\section{Quantum Langevin equations for the system}

We consider a driven optical cavity coupled by radiation pressure to a
micromechanical oscillator. The typical experimental configuration is a
Fabry-Perot cavity with one mirror much lighter than the other (see e.g.
\cite{karrai04,gigan06,arcizet06,arcizet06b,bouwm,harris}), but our
treatment applies to other configurations such as the silica toroidal
microcavity of Refs.~\cite{vahala1,vahalacool}. Radiation pressure typically
excites several mechanical degrees of freedom of the system with different
resonant frequencies. However, a single mechanical mode can be considered
when a bandpass filter in the detection scheme is used \cite{Pinard} and
coupling between the different vibrational modes can be neglected. The Hamiltonian of the system reads \cite%
{GIOV01}
\begin{eqnarray}
&& H=\hbar\omega_{c}a^{\dagger}a+\frac{1}{2}\hbar\omega_{m}(p^{2}+q^{2})-%
\hbar G_{0}a^{\dagger}a q  \notag \\
&& +i\hbar E(a^{\dagger}e^{-i\omega_{0}t}-ae^{i\omega_{0}t}).  \label{ham0}
\end{eqnarray}
The first term describes the energy of the cavity mode, with lowering
operator $a$ ($[a,a^{\dag}]=1$), cavity frequency $\omega_c$ and decay rate $%
\kappa$. The second term gives the energy of the mechanical mode, modeled as harmonic oscillator at frequency $\omega_m$ and
described by dimensionless position and momentum operators $q$ and $p$ ($%
[q,p]=i$). The third term is the radiation-pressure coupling of rate $%
G_0=(\omega_c/L)\sqrt{\hbar/m \omega_m}$, where $m$ is the effective mass of
the mechanical mode \cite{Pinard}, and $L$ is an effective length that
depends upon the cavity geometry: it coincides with the cavity length in the
Fabry-Perot case, and with the toroid radius in the case of Refs.~\cite%
{vahala1,vahalacool}. The last term describes the input driving by a laser with frequency $\omega_0$, where $E$ is related to the input laser
power $P$ by $|E|=\sqrt{2P \kappa/\hbar \omega_0}$. One can adopt the single cavity mode description of Eq.~(\ref{ham0}) as long as one drives
only one cavity mode and the mechanical frequency $\omega_m$ is much smaller than the cavity free spectral range $FSR \sim c/L$. In this case,
scattering of photons from the driven mode into other cavity modes is negligible \cite{law}.

The dynamics are also determined by the fluctuation-dissipation processes
affecting both the optical and the mechanical mode. They can be taken into
account in a fully consistent way \cite{GIOV01} by considering the following
set of nonlinear QLE, written in the interaction picture with respect to $%
\hbar \omega_0 a^{\dag}a$
\begin{subequations}
\label{nonlinlang}
\begin{eqnarray}
\dot{q}&=&\omega_m p, \\
\dot{p}&=&-\omega_m q - \gamma_m p + G_0 a^{\dag}a + \xi, \\
\dot{a}&=&-(\kappa+i\Delta_0)a +i G_0 a q +E+\sqrt{2\kappa} a^{in},
\end{eqnarray}
where $\Delta_0=\omega_c-\omega_0$. The mechanical mode is affected by a
viscous force with damping rate $\gamma_m$ and by a Brownian stochastic
force with zero mean value $\xi$, that obeys the correlation function \cite%
{Landau,GIOV01}
\end{subequations}
\begin{equation}  \label{browncorre}
\left \langle \xi(t) \xi(t^{\prime})\right \rangle = \frac{\gamma_m}{\omega_m%
} \int \frac{d\omega}{2\pi} e^{-i\omega(t-t^{\prime})} \omega \left[%
\coth\left(\frac{\hbar \omega}{2k_BT}\right)+1\right],
\end{equation}
where $k_B$ is the Boltzmann constant and $T$ is the temperature of the
reservoir of the micromechanical oscillator. The Brownian noise $\xi(t)$ is
a Gaussian quantum stochastic process and its non-Markovian nature (neither
its correlation function nor its commutator are proportional to a Dirac
delta) guarantees that the QLE of Eqs.~(\ref{nonlinlang}) preserve the
correct commutation relations between operators during the time evolution
\cite{GIOV01}. The cavity mode amplitude instead decays at the rate $\kappa$
and is affected by the vacuum radiation input noise $a^{in}(t)$, whose
correlation functions are given by
\begin{equation}
\langle a^{in}(t)a^{in,\dag}(t^{\prime})\rangle =\left[N(\omega_c)+1\right]%
\delta (t-t^{\prime}),  \label{input1}
\end{equation}
and
\begin{equation} \langle a^{in,\dag}(t)a^{in}(t^{\prime})\rangle =N(\omega_c)\delta (t-t^{\prime}), \label{input2}
\end{equation}
where $N(\omega_c)=\left(\exp\{\hbar \omega_c/k_BT\}-1\right)^{-1}$ is the
equilibrium mean thermal photon number. At optical frequencies $\hbar
\omega_c/k_BT \gg 1$ and therefore $N(\omega_c)\simeq 0$, so that only the
correlation function of Eq.~(\ref{input1}) is relevant.

Cooling of the mechanical oscillator by radiation pressure can be described in thermodynamical terms in the following way. Radiation pressure
couples the oscillator to the optical cavity mode, which behaves as an effective additional reservoir for the oscillator when the cavity is
appropriately detuned. As a consequence, the effective temperature of the mechanical mode will be intermediate between the initial reservoir
temperature and that of the effective optical
reservoir, which is in practice equal to zero due to the condition $%
N(\omega_c)\simeq 0$. Therefore one approaches the mechanical ground state when the coupling rate to the optical reservoir is much larger than
the damping rate $\gamma_m$ which gives the coupling to the initial reservoir. This explains why significant cooling is obtained when radiation
pressure coupling is strong. It is realized when the coupling $G_0$ is large, but is more easily achieved when the intracavity field is very
intense, i.e., for high-finesse cavities and enough driving power. In this limit (and if the system is stable) the system is characterized by a
semiclassical
steady state with the cavity mode in a coherent state with amplitude $%
\alpha_s$ ($|\alpha_s| \gg 1$), and a new equilibrium position for the
oscillator, displaced by $q_s$. The parameters $\alpha_s$ and $q_s$ are the
solutions of the nonlinear algebraic equations obtained by factorizing Eqs.~(%
\ref{nonlinlang}) and setting the time derivatives to zero. They are given
by
\begin{eqnarray}
&&q_s = \frac{G_0 |\alpha_s|^2}{\omega_m}, \\
&& \alpha_s = \frac{E}{\kappa+i \Delta},
\end{eqnarray}
where the latter equation is in fact the nonlinear equation determining $%
\alpha_s$, since the effective cavity detuning $\Delta$, including radiation
pressure effects, is given by
\begin{equation}
\Delta = \Delta_0- \frac{G_0^2 |\alpha_s|^2}{\omega_m}.
\end{equation}
Rewriting each Heisenberg operator of Eqs.~(\ref{nonlinlang}) as the
c-number steady state value plus an additional fluctuation operator with
zero mean value, one gets the exact QLE for the fluctuations
\begin{eqnarray}  \label{mode}
\delta \dot{q}&=&\omega_m \delta p,  \notag \\
\delta \dot{p}&=&-\omega_m \delta q - \gamma_m \delta p + G_0 \left(\alpha_s
\delta a^{\dag}+ \alpha_s ^* \delta a \right)+ \delta a^{\dag} \delta a +
\xi,  \notag \\
\delta \dot{a}&=&-(\kappa+i\Delta)\delta a +i G_0 \left(\alpha_s + \delta
a\right) \delta q +\sqrt{2\kappa} a^{in}.
\end{eqnarray}
We have assumed $|\alpha_s| \gg 1$, therefore one can safely neglect the nonlinear terms $\delta a^{\dag} \delta a$ and $\delta a \delta q$ in
the equations above and obtains the linearized QLE
\begin{subequations}
\label{lle}
\begin{eqnarray}
\delta \dot{q}&=&\omega_m \delta p, \\
\delta \dot{p}&=&-\omega_m \delta q - \gamma_m \delta p + G \delta X +\xi, \\
\delta \dot{X}&=&-\kappa \delta X+\Delta \delta Y +\sqrt{2\kappa} X^{in}, \\
\delta \dot{Y}&=&-\kappa \delta Y-\Delta \delta X +G\delta q +\sqrt{2\kappa}
Y^{in}.
\end{eqnarray}
Here we have chosen the phase reference of the cavity field so that $%
\alpha_s $ is real and positive, and we have defined the cavity field quadratures $\delta X\equiv(\delta a+\delta a^{\dag})/\sqrt{2}$ and
$\delta Y\equiv(\delta a-\delta a^{\dag})/i\sqrt{2}$, and the corresponding Hermitian
input noise operators $X^{in}\equiv(a^{in}+a^{in,\dag})/\sqrt{2}$ and $%
Y^{in}\equiv(a^{in}-a^{in,\dag})/i\sqrt{2}$. The linearized QLE show that
the mechanical mode is coupled to the cavity mode quadrature fluctuations by
the effective optomechanical coupling
\end{subequations}
\begin{equation}
G=G_0 \alpha_s\sqrt{2}=\frac{2\omega_c}{L}\sqrt{\frac{P \kappa}{m \omega_m
\omega_0 \left(\kappa^2+\Delta^2\right)}},  \label{optoc}
\end{equation}
which can be made very large by increasing the intracavity amplitude $%
\alpha_s$. Notice that together with the condition $\omega_m \ll c/L$ which
is required for the single cavity mode description, $|\alpha_s| \gg 1$ is
the \emph{only} assumption required by the present approach. We shall see
below that, thanks to this fact, the present approach provides a
generalization of previous treatments of back-action cooling, such as the
semiclassical treatment of Ref.~\cite{brag} and the perturbation treatments
of \cite{marquardt,wilsonrae}.

\section{Detuning-induced back-action cooling}

We have to evaluate the mean energy of the oscillator in the steady state
\begin{equation}
U=\frac{\hbar \omega _{m}}{2}\left[ \left\langle \delta q^{2}\right\rangle +\left\langle \delta p^{2}\right\rangle \right] \equiv \hbar \omega
_{m}\left( n_{eff}+\frac{1}{2}\right),   \label{meanener}
\end{equation}%
and to see if and when it approaches the ground-state value $%
\hbar \omega _{m}/2$. This is equivalent to determine the conditions
under which $ \left\langle \delta q^{2}\right\rangle \simeq
\left\langle \delta p^{2}\right\rangle \simeq 1/2$. The two
oscillator variances $\left\langle \delta q^{2}\right\rangle $ and
$\left\langle \delta p^{2}\right\rangle $ can be obtained by solving
Eqs.~(\ref{lle}) in the frequency domain and integrating the
corresponding fluctuation spectrum. One gets
\begin{equation}
\left\langle \delta q^{2}\right\rangle =\int_{-\infty }^{\infty }\frac{%
d\omega }{2\pi }S_{q}^{\Delta }(\omega ),\;\;\;\;\left\langle \delta
p^{2}\right\rangle =\int_{-\infty }^{\infty }\frac{d\omega }{2\pi }\frac{%
\omega ^{2}}{\omega _{m}^{2}}S_{q}^{\Delta }(\omega ),  \label{spectra}
\end{equation}%
where the position spectrum is given by
\begin{equation}
S_{q}^{\Delta }(\omega )=|\chi _{eff}^{\Delta }(\omega )\Vert ^{2}[S_{th}(\omega )+S_{rp}(\omega ,\Delta )],
\end{equation}%
where
\begin{equation}
S_{th}(\omega )=\frac{\gamma _{m}\omega }{\omega _{m}}\coth \left( \frac{%
\hbar \omega }{2k_{B}T}\right)   \label{spectratherm}
\end{equation}%
is the thermal noise spectrum,
\begin{equation}
S_{rp}(\omega ,\Delta )=\frac{G^{2}\kappa \left[ \Delta ^{2}+\kappa
^{2}+\omega ^{2}\right] }{\left[ \kappa ^{2}+(\omega -\Delta )^{2}\right] %
\left[ \kappa ^{2}+(\omega +\Delta )^{2}\right] }  \label{spectrarpn}
\end{equation}%
is the radiation pressure noise spectrum, and
\begin{equation}
\chi _{eff}^{\Delta }(\omega )=\omega _{m}\left[ \omega _{m}^{2}-\omega
^{2}-i\omega \gamma _{m}-\frac{G^{2}\Delta \omega _{m}}{(\kappa -i\omega
)^{2}+\Delta ^{2}}\right] ^{-1}  \label{chieffD}
\end{equation}%
is the effective susceptibility of the oscillator, modified by radiation pressure. The latter can be read as the susceptibility of an oscillator
with effective resonance frequency and damping rate given by
\begin{eqnarray}
&&\omega _{m}^{eff}(\omega )=\left[ \omega _{m}^{2}-\frac{G^{2}\Delta \omega
_{m}(\kappa ^{2}-\omega ^{2}+\Delta ^{2})}{\left[ \kappa ^{2}+(\omega
-\Delta )^{2}\right] \left[ \kappa ^{2}+(\omega +\Delta )^{2}\right] }\right]
^{\frac{1}{2}},  \label{omegeff} \\
&&\gamma _{m}^{eff}(\omega )=\gamma _{m}+\frac{2G^{2}\Delta \omega
_{m}\kappa }{\left[ \kappa ^{2}+(\omega -\Delta )^{2}\right] \left[
\kappa ^{2}+(\omega +\Delta )^{2}\right] } \label{dampeff}.
\end{eqnarray}%
\begin{figure}[tbh]
\centerline{\includegraphics[width=0.45\textwidth]{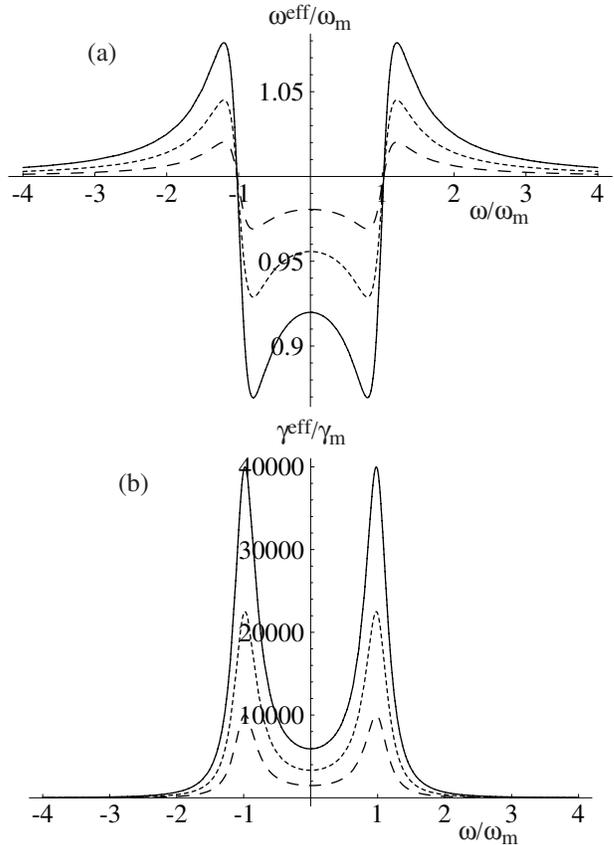}}
\caption{Plot of the effective mechanical frequency of
Eq.~(\protect\ref{omegeff}) (a) and of the effective mechanical
damping of Eq.~(\protect\ref{dampeff}) (b) versus frequency.
Parameter values are $\protect\omega _{m}/2\protect\pi =10$ MHz, $
\protect\gamma _{m}/2\protect\pi =100$ Hz, $\Delta =\protect\omega
_{m}$, $\protect\kappa =0.2$ $\protect\omega _{m}$ and $G=0.2$
$\protect\omega _{m}$ (dashed line), $G=0.3$ $\protect\omega _{m}$
(dotted line) and $G=0.4$ $\protect\omega _{m} $ (full line), which
we shall see later correspond to an optimal cooling regime.}
\label{Fig1}
\end{figure}
The modification of the mechanical frequency due to radiation pressure shown by Eq.~(\ref{omegeff}) is the so-called ``optical spring effect'',
which may lead to significant frequency shifts in the case of low-frequency oscillators, such as pendulum modes of suspended mirrors
\cite{mavalvala}. In the case of higher resonance frequencies, such as those of Refs.~\cite{gigan06,arcizet06b,vahalacool} where $\omega_m
\gtrsim 1$ MHz, the optical spring term in Eq.~(\ref{omegeff}) does not significantly alter the frequency, even for large intracavity power.
Here, we shall only consider numerical examples with large $\omega_m$, where the frequency is practically unchanged ($\omega _{m}^{eff}(\omega
)\simeq \omega _{m}$, see Fig.~\ref{Fig1}a). In fact, ground state cooling $n_{eff}< 1$ can be approached only if the initial mean thermal
excitation number $\bar{n}=\left(\exp\{\hbar \omega_m/k_BT\}-1\right)^{-1}$ is not prohibitively large, and this is possible, even at cryogenic
temperatures, only if $\omega_m$ is sufficiently large. For positive $\Delta $ and for large enough $G$ the effective mechanical damping is
instead significantly increased (see Fig.~\ref{Fig1}b). This increase is at the basis of the cooling process. In fact, the mechanical
susceptibility at resonance is inversely proportional to damping and it is therefore significantly suppressed by radiation pressure. As a
consequence, the oscillator is much less affected by thermal noise and this means cooling as long as the radiation-pressure noise contribution
$S_{rp}(\omega ,\Delta )$ remains small compared to thermal noise term $S_{th}(\omega)$, which is verified for not too large $G$.

Let us now determine the oscillator mean energy. The system reaches a steady state only if it is stable and this is satisfied when all the poles
of the effective susceptibility $\chi_{eff}^{\Delta}$ lie in the lower complex half-plane. By applying the Routh-Hurwitz criterion \cite{grad},
we get the following two nontrivial stability conditions in the detuned-cavity case:
\begin{subequations}
\label{stab}
\begin{eqnarray}
&& s_1=2\gamma_m\kappa\left\{ \left[ \kappa^{2}+\left(
\omega_m-\Delta\right) ^{2}\right] \left[ \kappa^{2}+\left(
\omega_m+\Delta\right) ^{2}\right] \right.  \notag \\
&&\left.+\gamma_m\left[ \left( \gamma_m+2\kappa\right) \left(
\kappa^{2}+\Delta ^{2}\right) +2\kappa\omega_{m}^{2}\right] \right\}  \notag
\\
&&+\Delta\omega_{m} G^{2}\left( \gamma_m+2\kappa\right) ^{2}>0, \\
&&s_2=\omega_{m}\left( \kappa^{2}+\Delta^{2}\right) -G^{2}\Delta>0.
\end{eqnarray}
When the stability conditions are satisfied, the integrals of Eq.~(\ref{spectra}) for the two variances can be solved exactly. However, it is
reasonable to simplify the thermal noise contribution in Eqs.~(\ref{spectratherm}). In fact $k_B T/\hbar \simeq 10^{11}$ s$^{-1}$ even at
cryogenic temperatures and therefore is always much larger than all the other parameters. At these high values of $\omega$ the position spectrum
is negligible and therefore one can safely approximate in the integral
\end{subequations}
\begin{equation}  \label{thermappr}
\frac{\gamma_m \omega}{\omega_m} \coth\left(\frac{\hbar \omega}{2k_BT}%
\right) \simeq \gamma_m \frac{2k_B T}{\hbar \omega_m} \simeq \gamma_m\left(2%
\bar{n}+1\right).
\end{equation}

Performing the integrals, one gets the final expressions for the two variances, which are given by

\begin{eqnarray}
\langle \delta q^2\rangle &=& \frac{1}{2}+b_q+d_q \bar{n}, \label{qrel-correct} \\
\langle \delta p^2\rangle &=& \frac{1}{2}+b_p+d_p \bar{n}, \label{prel-correct}
\end{eqnarray}
where
\begin{widetext}
\begin{eqnarray}
b_p & =& [s_1]^{-1}G^2 \kappa \left\{\Delta^2(\gamma_m+\kappa)+\kappa (\gamma_m \kappa +\kappa^2+\omega_m^2)-\Delta\omega_m (\gamma_m+2\kappa)
\right\}, \label{bip}\\
d_p & =& 1-[s_1]^{-1} 2 G^2 \kappa \omega_m \Delta (\gamma_m+2\kappa), \label{dip}\\
b_q & =& [2s_1s_2]^{-1}G^2 \left\{ 2 \kappa \left( \Delta^2 + \kappa^2 \right) \left\{ \left[ \Delta^2 + {\left( \gamma_m + \kappa \right) }^2
\right]\left(\kappa\omega_m+\gamma_m
\Delta\right)\right. \right.\label{biq}\\
&+ & \left. \left. \omega_m^2\left(\gamma_m+\kappa\right)\left(\omega_m-2\Delta\right)\right\}+
\Delta G^2\omega_m\left(\gamma_m+2\kappa\right)\left[\Delta \gamma_m-\kappa \left(\omega_m-2\Delta\right)\right]\right\}, \nonumber \\
d_q & =& 1+[s_1s_2]^{-1}\Delta G^2\left[s_1-2\gamma_m\kappa \omega_m^2\left(\omega_m^2+2\gamma_m \kappa +4 \kappa^2\right)- 4\kappa^2
\omega_m^2\left(\Delta^2+\kappa^2\right)\right]. \label{diq}
\end{eqnarray}
\end{widetext}
At the ground state both variances are equal to $1/2$ and therefore realizing ground state cooling means achieving $b_q$, $b_p$, $d_q$, $d_p \to
0$. Notice that in general $ \left\langle \delta q^{2}\right\rangle \neq \left\langle \delta p^{2}\right\rangle$, that is, one does not have
energy equipartition, as it is already shown by the general expression of Eq.~(\ref{spectra}). This means that in the generic case, the steady
state of the system is not, strictly speaking, a thermal equilibrium state and this prevents to derive from here an univocally defined
temperature. With this respect, Eq.~(\ref{meanener}) provides a definition of the effective mean excitation number $n_{eff}$ from which one can
only define an effective temperature as $T_{eff}=\hbar \omega_m/\left[k_B \ln\left(1+1/n_{eff}\right)\right]$. However, in order to get to the
quantum ground state, both variances have to tend to $1/2$ and therefore energy equipartition has to be satisfied in the optimal regime close to
the ground state.

In order to have an intuitive picture and to facilitate the comparison with the recent perturbation treatments of
Refs.~\cite{marquardt,wilsonrae}, we consider the expressions of the variances in some limiting case of experimental interest.

It is convenient to introduce the rates~\cite{wilsonrae}
\begin{equation}
A_{\pm }=\frac{G^{2}\kappa }{2\left[ \kappa ^{2}+\left( \Delta \pm \omega
_{m}\right) ^{2}\right] },
\end{equation}%
which define the rates at which laser photons are scattered by the moving oscillator simultaneously with the absorption (Stokes, $A_{+}$) or
emission (anti-Stokes, $A_{-}$) of the oscillator vibrational phonons. For $\Delta >0$ one has $A_{-}>A_{+}$ and a net laser cooling rate
$$\Gamma =A_{-}-A_{+}>0$$
can be defined, giving the rate at which mechanical energy is taken away by the leaking cavity. As a consequence, the total mechanical damping
rate is given by $\gamma_m+\Gamma$, which is consistent with the expression of the effective (frequency-dependent) damping rate of
Eq.~(\ref{dampeff}): in fact, it is easy to check that \begin{equation} \gamma _{m}^{eff}(\omega=\omega_m )=\gamma_m+\Gamma .\label{gammeff2}
\end{equation}
As discussed in the preceding section, ground-state cooling is achievable when $\gamma _{m}$, the coupling rate with the thermal reservoir, is
significantly smaller than $\Gamma $, which represents the coupling rate of the mechanical oscillator with the effective reservoir provided by
the damped cavity mode. For small $\gamma _{m}$ the above expressions simplify to
\begin{equation}
\left\langle \delta p^{2}\right\rangle =\frac{1}{\gamma _{m}+\Gamma }\left\{
\frac{A_{+}+A_{-}}{2}+\gamma _{m}\bar{n}\left( 1+\frac{\Gamma }{2\kappa }%
\right) \right\} ,  \label{pisq}
\end{equation}%
\begin{equation}
\left\langle \delta q^{2}\right\rangle =\frac{1}{\gamma _{m}+\Gamma }\left\{
a\frac{A_{+}+A_{-}}{2}+\frac{\gamma _{m}\bar{n}}{\eta _{\Delta }}\left( 1+%
\frac{\Gamma }{2\kappa }b\right) \right\} ,  \label{qusq}
\end{equation}%
where we have defined the coefficients
\begin{eqnarray}
a &=&\frac{\kappa ^{2}+\Delta ^{2}+\eta _{\Delta }\omega _{m}^{2}}{\eta
_{\Delta }\left( \kappa ^{2}+\Delta ^{2}+\omega _{m}^{2}\right) }, \\
b &=&\frac{2\left( \Delta ^{2}-\kappa ^{2}\right) -\omega _{m}^{2}}{\kappa
^{2}+\Delta ^{2}}, \\
\eta _{\Delta } &=&1-\frac{G^{2}\Delta }{\omega _{m}(\kappa ^{2}+\Delta ^{2})%
}.
\end{eqnarray}%
Note that for positive $\Delta $, $0<\eta _{\Delta }<1$ due to the stability condition of Eq.~(\ref{stab}b). Eqs.~(\ref{pisq}) and (\ref{qusq})
provide a generalization of the results of Ref.~\cite{marquardt,wilsonrae}, which are reproduced if we take $\omega _{m}\gg \bar{n}\gamma
_{m},G$ and $\kappa \gg
\gamma _{m},G$ (assumed in Refs.~\cite{marquardt,wilsonrae}) in Eqs.~(\ref%
{pisq}) and (\ref{qusq}). In these limits $a,\eta _{\Delta }\rightarrow 1$, $%
\Gamma /\kappa \rightarrow 0$ and therefore
\begin{equation}
\left\langle \delta p^{2}\right\rangle \simeq \left\langle \delta
q^{2}\right\rangle =n_{eff}+1/2,
\end{equation}%
where
\begin{equation}
n_{eff}=\frac{\gamma _{m}\bar{n}+A_{+}}{\gamma _{m}+\Gamma }  \label{neffdet}
\end{equation}%
is the basic result of Ref.~\cite{marquardt} (see Eq.~(9)) and Ref.~\cite%
{wilsonrae} (see Eq.~(5) and its derivation). This result has been also obtained in Ref.~\cite{dantan} through an approximate treatment of the
exact integrals of Eq.~(\ref{spectra}).

\begin{figure}[tbh]
\centerline{\includegraphics[width=0.45\textwidth]{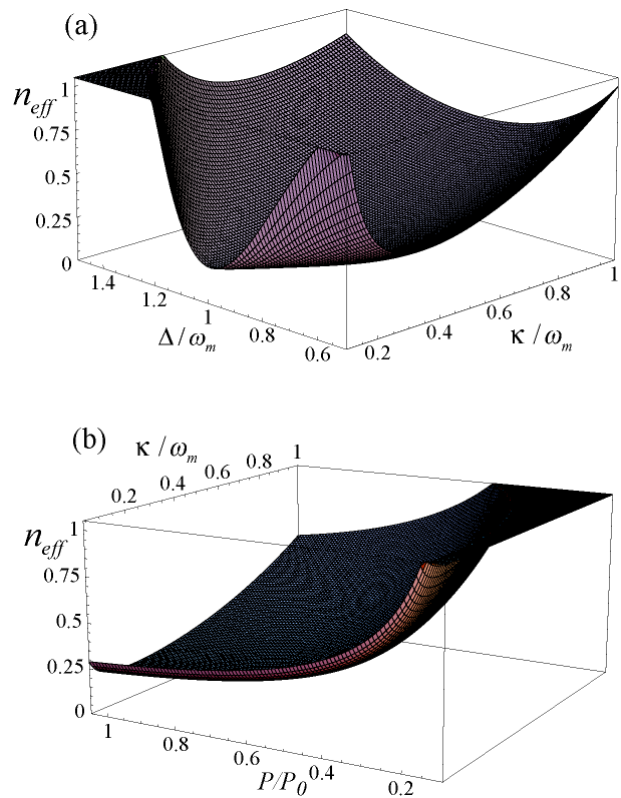}} \caption{(Color online) a) Effective mean vibrational number $n_{eff}$
versus
$\Delta /\protect\omega _{m}$ and $\kappa /\protect\omega _{m}$ around the optimal ground-state cooling regime for $\protect%
\omega _{m}/2\protect\pi =10$ MHz, $\protect\gamma _{m}/2\protect\pi =100$ Hz, $m=250$ ng, a cavity of length $L=0.5$ mm driven by a laser with
power $P=50 $ mW, and wavelength $1064$ nm. The oscillator reservoir temperature is $T=0.6$ K, corresponding to $\bar{n}\simeq 1250$. The
minimum value $n_{eff}\simeq 0.1$ corresponds to an effective temperature $T_{eff}\simeq 0.2$ mK. (b) Effective mean vibrational number
$n_{eff}$ versus $\kappa /\protect\omega _{m}$ and the normalized power $P/P_0$, ($P_0=50$ mW) at the fixed, optimal value for the detuning, $
\Delta =\protect\omega _{m}$. The other parameters are the same as in (a).} \label{Fig2}
\end{figure}

The approximate expression of Eq.~(\ref{neffdet}) suggests how the system parameters can be chosen in order to minimize the oscillator energy
close to the ground-state value. In order to reach the condition $n_{eff}<1$ one needs a large $\Gamma/\gamma_m $ while keeping $A_{+}/\Gamma$
small. This is done by matching $\Delta =\omega_m$, which optimizes the energy transfer from the mechanical mode to the anti-Stokes sideband.
Further optimization requires a large value of $G$, which is however constrained by the stability condition $\eta_{\Delta}
> 0$. When $\eta_{\Delta} \to 0$, $\left\langle \delta q^{2}\right\rangle$ becomes too large, while the ground state, where $\left\langle \delta
q^{2}\right\rangle=\left\langle \delta p^{2}\right\rangle=1/2$, is approached for $\eta _{\Delta } \sim 1$ (see Eq.~(\ref{qusq})). Finally the
optimal ratio $\kappa /\omega _{m}$ is determined by the fact that the laser-cooling rate $\Gamma$ has to be large but still smaller than
$\kappa$  because the cavity response time $\sim \kappa^{-1}$ has to be remain shorter than that of the oscillator $\sim \Gamma^{-1}$. We find
that, within a set of experimentally achievable parameters, the best cooling regime ($ n_{eff}\simeq 0.1$), is obtained in the good cavity limit
condition $\kappa /\omega _{m}\simeq 0.2$ (see Fig.~\ref{Fig2}), which is close to the value of $1/ \sqrt{32}$ suggested in \cite{wilsonrae}.

\section{Ground-state cooling with cold damping}

An alternative way of cooling the oscillator by overdamping, proposed in
\cite{Mancini98} and experimentally realized in \cite%
{cohadon99,arcizet06,bouwm,rugar}, is to use quantum feedback and specifically cold damping \cite{courty,quiescence02,zolingen}. This technique
is based on the application of a negative derivative feedback, which increases the damping of the system without increasing the thermal noise.
The oscillator position is measured by means of a phase-sensitive detection of the cavity output, which is then fed back to the oscillator by
applying a force whose intensity is proportional to the time derivative of the output signal, and therefore to the oscillator velocity. One
measures
the phase quadrature $Y$, whose Fourier transform, according to Eqs.~(\ref%
{lle}), is given by
\begin{equation}  \label{fty}
\delta Y(\omega)=\frac{G (\kappa - i\omega)}{(\kappa - i\omega)^2+\Delta^2}%
\delta q(\omega)+\text{noise terms}.
\end{equation}
This shows that the highest sensitivity for position measurements is achieved for a resonant cavity, $\Delta=0$ and in the large cavity
bandwidth limit $\kappa \gg \omega_m, \gamma_m$, i.e., when the cavity mode adiabatically follows the oscillator dynamics $\delta
Y(\omega)\simeq (G/\kappa)\delta q(\omega)$. Therefore the QLE for the cold-damping scheme coincide with those of Eqs.~(\ref{lle}) with
$\Delta=0$, but with an additional feedback force
\begin{subequations}
\label{llecd}
\begin{eqnarray}
\delta \dot{q}&=&\omega_m \delta p, \\
\delta \dot{p}&=&-\omega_m \delta q - \gamma_m \delta p + G \delta X +\xi, \\
&&- \int_{-\infty}^{t} ds g(t-s)\delta Y^{est}(s),  \notag \\
\delta \dot{X}&=&-\kappa \delta X +\sqrt{2\kappa} X^{in}, \\
\delta \dot{Y}&=&-\kappa \delta Y +G\delta q +\sqrt{2\kappa} Y^{in}.
\end{eqnarray}
\end{subequations}
In Eq.~(\ref{llecd}b) $g(t)$ is a causal kernel, proportional to a derivative of a Dirac delta in the ideal derivative feedback limit, and
$\delta Y^{est}(s)$ is the estimated intra-cavity phase quadrature that is obtained from the measurement of the output quadrature $Y^{out}(t)$
as follows. The usual input-output relation
\begin{equation}  \label{inout0}
\delta Y^{out}(t)= \sqrt{2\kappa}\delta Y(t)- Y^{in}(t)
\end{equation}
can be generalized to the case of a non-unit detection efficiency, by modeling a detector with quantum efficiency $\eta$ with an ideal detector
preceded by a beam splitter with transmissivity $\sqrt{\eta}$, hence mixing the incident field with an uncorrelated vacuum field $Y^v(t)$
\cite{gard}. The generalized input-output relation then reads
\begin{equation}  \label{inout}
Y^{out}(t)= \sqrt{\eta}\left[\sqrt{2\kappa}\delta Y(t)-  Y^{in}(t)\right]-\sqrt{1-\eta}Y^v(t),
\end{equation}
so that the estimated phase quadrature $\delta Y^{est}(s)$ is given by \cite{footnote}
\begin{equation}  \label{inoutest} \delta Y^{est}(t)\equiv \frac{Y^{out}(t)}{\sqrt{2\eta \kappa}}=
\delta Y(t)-\frac{ Y^{in}(t)+\sqrt{\eta^{-1}-1}Y^v(t)}{\sqrt{2\kappa}}.
\end{equation}
After solving the QLE of Eqs.~(\ref{llecd}) by Fourier transform, one finds that the two oscillator variances in the cold-damping case are given
again by
Eqs.~(\ref{spectra}), but with a different position spectrum $%
S_{q}^{cd}(\omega )$, that can be expressed as
\begin{equation}
S_{q}^{cd}(\omega )=|\chi _{eff}^{cd}(\omega )\Vert ^{2}[S_{th}(\omega )+S_{rp}(\omega ,0)+S_{fb}(\omega )].  \label{spectraeffcd}
\end{equation}%
$S_{th}(\omega )$ is given by Eq.~(\ref{spectratherm}), $S_{rp}(\omega ,0)$ is obtained by choosing $\Delta =0$ in Eq.~(\ref{spectrarpn}), and
one has the additional contribution
\begin{equation}
S_{fb}(\omega )=\frac{|g(\omega )|^{2}}{4\kappa \eta }  \label{spectrafb}
\end{equation}%
due to the measurement noise that is fed back into the oscillator dynamics by the cold-damping feedback loop ($g(\omega )$ is the Fourier
transform of $g(t)$). Finally,
\begin{equation}
\chi _{eff}^{cd}(\omega )=\omega _{m}\left[ \omega _{m}^{2}-\omega
^{2}-i\omega \gamma _{m}+\frac{g(\omega )G\omega _{m}}{\kappa -i\omega }%
\right] ^{-1}  \label{chieffcd}
\end{equation}%
is the effective susceptibility of the oscillator in the cold-damping scheme, which depends upon the explicit form of the feedback transfer
function $g(\omega )$. The simplest choice, corresponding to a standard derivative high-pass filter, is
\begin{equation}
g(\omega )=\frac{-i\omega g_{cd}}{1-i\omega /\omega _{fb}},
\label{feedtrans}
\end{equation}%
which means choosing
\begin{equation*}
g(t)=g_{cd}\frac{d}{dt}\left[ \theta (t)\omega _{fb}e^{-\omega _{fb}t}\right]
\end{equation*}%
so that $\omega _{fb}^{-1}$ plays the role of the time delay of the feedback loop, and $g_{cd}>0$ is the feedback gain. The ideal derivative
limit is obtained for $\omega _{fb}\rightarrow \infty $, implying $g(\omega )=-i\omega g_{cd}$ and therefore $g(t)=g_{cd}\delta ^{\prime }(t)$.
The cold-damping susceptibility of Eq.~(\ref{chieffcd}) can be read again as the susceptibility of an oscillator with effective resonance
frequency and damping rate given by

\begin{figure}[tbh]
\centerline{\includegraphics[width=0.45\textwidth]{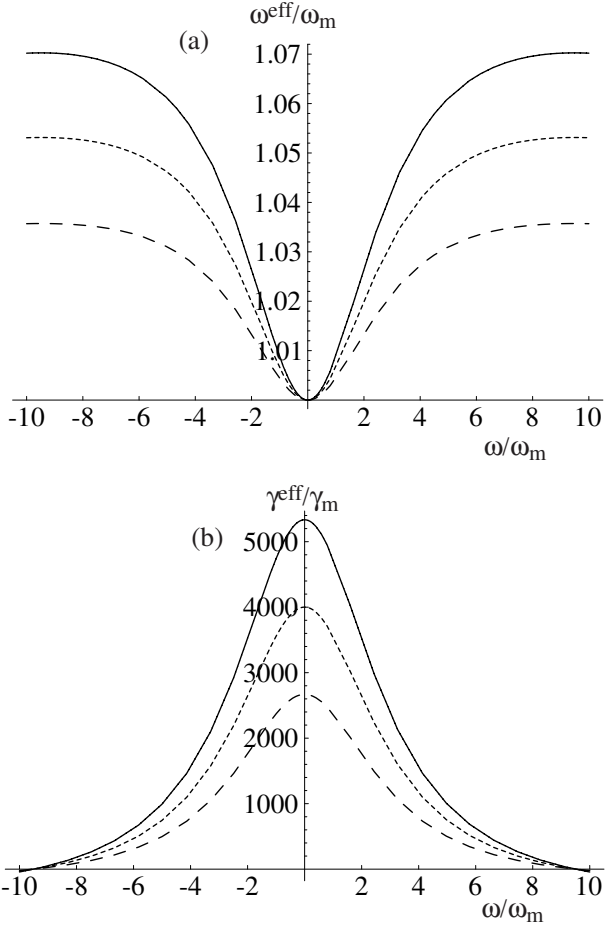}}
\caption{Plot of the effective mechanical frequency of
Eq.~(\protect\ref{omegeffcd}) (a), and of the effective mechanical
damping of Eq.~(\protect\ref{dampeffcd}) (b), versus frequency.
Parameter values are $\protect\omega _{m}/2\protect\pi =10$ MHz, $
\protect\gamma _{m}/2\protect\pi =100$ Hz, $g_{cd}=4$,
$\omega_{fb}=3\omega _{m}$, $\kappa = 30 \omega _{m}$ and $G=0.2$
$\protect\omega _{m}$ (dashed line), $G=0.3$ $\protect\omega _{m}$
(dotted line) and $G=0.4$ $\protect\omega _{m} $ (full line), which
we shall see later correspond to an optimal feedback cooling
regime.} \label{Fig3}
\end{figure}

\begin{eqnarray}
&&\omega _{m}^{eff,cd}(\omega )=\left[ \omega _{m}^{2}+\frac{g_{cd}G\omega
_{m}\omega _{fb}\omega ^{2}(\kappa +\omega _{fb})}{(\kappa ^{2}+\omega
^{2})(\omega _{fb}^{2}+\omega ^{2})}\right] ^{\frac{1}{2}},
\label{omegeffcd} \\
&&\gamma _{m}^{eff,cd}(\omega )=\gamma _{m}+\frac{g_{cd}G\omega
_{m}\omega _{fb}(\kappa \omega _{fb}-\omega ^{2})}{(\kappa
^{2}+\omega ^{2})(\omega _{fb}^{2}+\omega ^{2})}.\label{dampeffcd}
\end{eqnarray}%
The frequency dependence of the effective resonance frequency and damping depends upon the specific form of the transfer function $g(\omega )$,
and the one associated with the choice of Eq.~(\ref{feedtrans}) is plotted in Fig.~3 for comparison with the corresponding curves for the
back-action cooling of Fig.~1. Fig.~3a shows again that in the chosen parameter regime, the frequency shift, i.e., the optical spring effect, is
negligible. Cold damping is usually applied in the adiabatic limit when $\kappa ,\omega _{fb}\gg \omega $ and in this limit one has $\omega
_{m}^{eff,cd}\simeq \omega _{m}$ and $\gamma _{m}^{eff,cd}\simeq \gamma _{m}+g_{cd}G\omega _{m}/\kappa =\gamma _{m}(1+g_{2})$, where we have
defined the scaled, dimensionless feedback gain $g_{2}\equiv g_{cd}G\omega _{m}/\kappa \gamma _{m}$ \cite{quiescence02}. In this limit the only
effect of cold damping is to increase the mechanical damping rate without significantly changing the resonance frequency
As a consequence, the system is stable whenever $g_{cd}\geq 0$. This is not any longer true in the general case when finite
values of $\kappa $ and $\omega _{fb}$ are considered. By imposing that all the poles of $\chi _{eff}^{cd}(\omega )$ lie in the lower complex
half-plane we get one nontrivial stability condition
\begin{eqnarray}
&&s_{cd}=\left[ \gamma _{m}\kappa \omega _{fb}+g_{cd}G\omega _{m}\omega
_{fb}+\omega _{m}^{2}(\kappa +\omega _{fb})\right]   \notag \\
&&\times \left[ (\kappa +\gamma _{m})(\kappa +\omega _{fb})(\gamma
_{m}+\omega _{fb})+\gamma _{m}\omega _{m}^{2}-g_{cd}G\omega _{m}\omega _{fb}%
\right]   \notag \\
&&-\kappa \omega _{m}^{2}\omega _{fb}(\kappa +\gamma _{m}+\omega
_{fb})^{2}>0.  \label{stabcd}
\end{eqnarray}%
This stability condition is always satisfied in the ideal derivative feedback limit $(\omega _{fb}\rightarrow \infty )$ and in the ideal
adiabatic limit $\kappa \rightarrow \infty $. However, for finite values of $%
\omega _{fb}$ and $\kappa $, Eq.~(\ref{stabcd}) implies an upper limit to
the feedback gain which is given by
\begin{equation}
g_{2}<\epsilon _{0}+\sqrt{\epsilon _{0}^{2}+\epsilon _{r}^{2}},
\label{g2max}
\end{equation}%
where
\begin{eqnarray}
\epsilon _{0} &=&\frac{1}{2}\left[ \frac{\omega _{fb}}{\gamma _{m}}\left( 1+%
\frac{\gamma _{m}}{\kappa }\right) +\frac{\kappa }{\gamma _{m}}+\frac{\gamma
_{m}}{\kappa }+1-\frac{\omega _{m}^{2}}{\kappa \gamma _{m}}\right.   \notag
\\
&&\left. +\frac{\omega _{m}^{2}}{\kappa \omega _{fb}}+\frac{\gamma
_{m}+\kappa }{\omega _{fb}}-\frac{\omega _{m}^{2}}{\gamma _{m}\omega _{fb}}%
\right] , \\
\epsilon _{r}^{2} &=&\frac{\omega _{m}^{2}+\kappa ^{2}+\gamma _{m}\kappa }{%
\omega _{fb}^{2}\kappa ^{2}\gamma _{m}}\left[ \omega _{fb}^{3}+\omega
_{fb}^{2}\left( \kappa +\gamma _{m}\right) \right.   \notag \\
&&\left. +\omega _{fb}\left( \omega _{m}^{2}+\gamma _{m}\kappa \right)
+\kappa \omega _{m}^{2}\right] .
\end{eqnarray}%
The system may become unstable for large gain because, for non-zero time delay, the feedback force can be out-of-phase with the oscillator
motion and become an accelerating rather than a viscous force.

The exact expression for the position and momentum variances can be obtained by integrating Eq.~(\ref{spectra}) using the corresponding spectrum
for the cold-damping case, given by Eq.~(\ref{spectraeffcd}). Using again the approximation of Eq.~(\ref{thermappr}), one gets
\begin{eqnarray}  \label{qusqcdgen}
\left\langle \delta q ^2\right\rangle & = & s_{cd}^{-1}\left\{\left(\bar{n}+
\frac{1}{2}\right)\left[A_{cd}+\left(1+\frac{\omega_{fb}^2}{\kappa^2}%
\right)B_{cd}+C_{cd}\right]\right.  \notag \\
&&\left.+\frac{g_{cd}^2\omega_{fb}^2}{8\kappa\eta} \left[A_{cd}+B_{cd}\right]%
+\frac{G^2}{2\kappa}\left[B_{cd}+C_{cd}\right]\right\}, \\
\left\langle \delta p ^2\right\rangle &=& s_{cd}^{-1}\left\{\left(\bar{n}+
\frac{1}{2}\right)\left[\left(\frac{\kappa^2+\omega_{fb}^2}{\omega_m^2}%
\right)A_{cd}\right.\right.  \label{pisqcdgen} \\
&&+\left.\left.\frac{\omega_{fb}^2}{\omega_m^2}B_{cd}+D_{cd}\right] +\frac{%
g_{cd}^2\omega_{fb}^2}{8\kappa\eta} \left[\frac{\kappa^2}{\omega_m^2}%
A_{cd}+D_{cd}\right]\right.  \notag \\
&&\left.+\frac{G^2}{2\kappa}\left[\frac{\kappa^2}{\omega_m^2}A_{cd}+\frac{%
\omega_{fb}^2}{\omega_m^2}B_{cd} \right]\right\},  \notag
\end{eqnarray}
where
\begin{eqnarray}
A_{cd} &=& \omega_m^2\left[\kappa\omega_m^2+\omega_{fb}\left(\omega_m^2+%
\gamma_m\kappa+g_{cd}G\omega_m\right)\right], \\
B_{cd} &=& \omega_m^2\kappa^2\left(\gamma_m+\kappa+\omega_{fb}\right), \\
C_{cd} &=& \omega_{fb}\kappa\left[\omega_{fb}^2\left(\kappa+\gamma_m\right)+%
\omega_{fb}\left(\kappa+\gamma_m\right)^2\right.  \notag \\
&+&\left.\gamma_m\left(\omega_m^2+\kappa^2+\kappa\gamma_m\right)\right], \\
D_{cd} &=& \omega_{fb}^2\left[\gamma_m\left(\omega_m^2+\kappa^2+\kappa%
\gamma_m\right)+\left(\kappa+\gamma_m\right)g_{cd}G\omega_m\right]  \notag \\
&+&\omega_{fb}\left(\omega_m^2+\kappa\gamma_m\right)\left(\omega_m^2+
\kappa\gamma_m+g_{cd}G\omega_m\right)  \notag \\
&+&\kappa\omega_m^2\left(\omega_m^2+\gamma_m\kappa\right).
\end{eqnarray}
Eqs.~(\ref{qusqcdgen}) and (\ref{pisqcdgen}) show that also with cold damping $\left\langle \delta q ^2\right\rangle \neq \left\langle \delta p
^2\right\rangle$, i.e., energy equipartition does not hold in general.

The optimal cooling conditions in the cold-damping scheme can be obtained by
minimizing the sum of the variances of Eqs.~(\ref{qusqcdgen}) and (\ref%
{pisqcdgen}), which is non trivial in general. Since, however, cold-damping feedback is designed to work only within the adiabatic limit, we can
restrict the discussion to the bad cavity limit where $\kappa \gg \omega_m,\gamma_m$. In fact, the feedback force is an additional viscous force
that is able to overdamp the mechanical oscillator only when the output signal
$\delta Y^{est}(t)$ is proportional to the oscillator position $\delta q(t)$%
, which happens when $\kappa$ is much larger than the relevant frequencies $%
\omega$ of the mechanical motion. In the good cavity limit ($\kappa \ll \omega_m,\gamma_m$) on the contrary, the output signal $\delta
Y^{est}(t)$ is proportional to the \emph{time integral} of the oscillator position $\delta q(t)$ and therefore the feedback force is
proportional to the oscillator position rather than to its velocity. This means that in the good cavity limit the feedback loop has no
cold-damping effect, because it increases the
mechanical frequency, $\omega_m^{eff,cd} \simeq \left[\omega_m^2+g_{cd}G%
\omega_m\right]^{1/2}$, without appreciably modifying the mechanical damping
(see Eqs.~(\ref{chieffcd}), (\ref{omegeffcd}), and (\ref{dampeffcd})).

We discuss the expressions of $\left\langle \delta q ^2\right\rangle$ and $\left\langle \delta p^2\right\rangle$ in the adiabatic limit by
distinguishing two situations that depend upon the value
of the feedback bandwidth $\omega_{fb}$: i) very large bandwidth, $%
\omega_{fb} \gg \kappa \gg \omega_m,\gamma_m$, where the feedback is practically instantaneous, ; ii) finite bandwidth, $\kappa \gg \omega_{fb}
\sim \omega_m \gg \gamma_m$. In the first case one has
\begin{eqnarray}  \label{qusqcd1}
\left\langle \delta q ^2\right\rangle & \simeq & \frac{\bar{n}+ \frac{1}{2}+%
\frac{\zeta}{4} +\frac{g_{2}^{2}}{4\eta\zeta }\left( 1+g_2 \frac{\gamma_m}{%
\kappa}\right)}{1+g_{2}} , \\
\left\langle \delta p ^2\right\rangle & \simeq & \frac{\left(\bar{n}+ \frac{1%
}{2}\right)\left( 1+g_2 \frac{\gamma_m}{\kappa}\right)+\frac{\zeta}{4}}{%
1+g_{2}}+\frac{g_{2}^{2}}{4\eta\zeta }\frac{\omega_{fb}\gamma_m}{\omega_m^2},
\label{pisqcd1}
\end{eqnarray}
where we have defined the scaled dimensionless input power $%
\zeta=2G^{2}/\kappa\gamma_m$. These results provide the generalization of the results of \cite{quiescence02,courty}, where the quantum limits of
cold-damping have been already discussed within the adiabatic limit. In fact,
Eqs.~(\ref{qusqcd1})-(\ref{pisqcd1}) reproduce the results of Ref.~\cite%
{quiescence02} in the large-bandwidth limit of the feedback except for the addition
of the non-adiabatic correction term $g_2 \gamma_m/\kappa$ for both $%
\left\langle \delta q ^2\right\rangle$ and $\left\langle \delta p
^2\right\rangle$. The almost instantaneous feedback regime $\omega_{fb} \gg
\kappa \gg \omega_m,\gamma_m$ is not convenient for cooling because of the
last contribution to $\left\langle \delta p ^2\right\rangle$, which is very
large since it diverges linearly with $\omega_{fb}$. This is due to the fact
that the derivative feedback injects a large amount of shot noise when its
bandwidth is very large.

\begin{figure}[tbh]
\centerline{\includegraphics[width=0.45\textwidth]{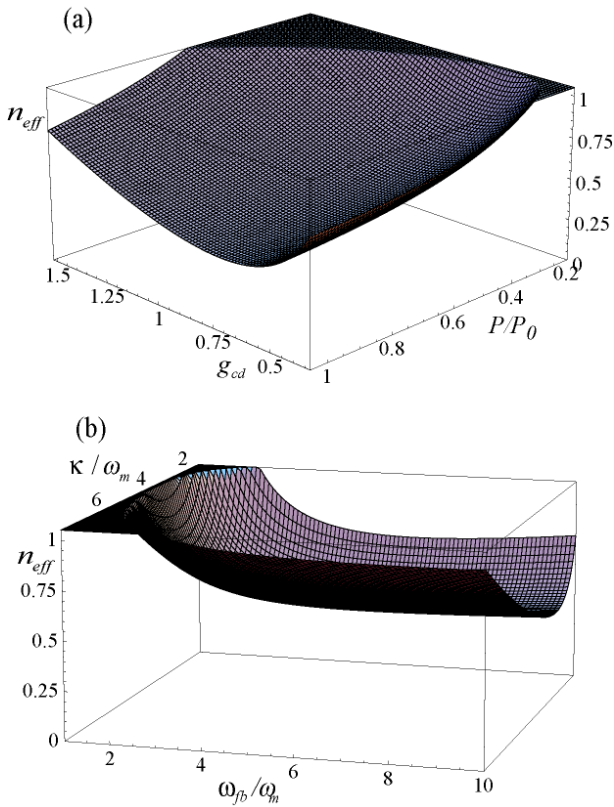}} \caption{(Color online) a) Effective mean vibrational number $n_{eff}$
versus the feedback gain $g_{cd}$ and the scaled input power $P/P_0$ ($P_0=100$ mW), around the optimal cooling regime for $\protect%
\omega _{m}/2\protect\pi =10$ MHz, $\protect\gamma _{m}/2\protect\pi =100$ Hz, $m=250$ ng, a cavity of length $L=0.5$ mm driven by a laser with
wavelength $1064$ nm and bandwidth $\kappa=3 \omega_m$. The feedback bandwidth is $\omega_{fb}=3.5 \omega_m$. The oscillator reservoir
temperature is $T=0.6$ K, corresponding to $\bar{n}\simeq 1250$. The minimum value $n_{eff}\simeq 0.2$ corresponds to an effective temperature
$T_{eff}\simeq 0.27$ mK. (b) Effective mean vibrational number $n_{eff}$ versus $\protect\kappa / \protect\omega _{m}$ and $\protect\omega
_{fb}/\protect\omega _{m}$ at fixed input power $P_0=100$ mW and feedback gain $g_{cd}=0.8$. The other parameters are as in (a).} \label{Fig4}
\end{figure}

In the other limit where the feedback delay time is comparable to the
oscillator timescales, that is, $\kappa \gg \omega _{fb}\sim \omega _{m}\gg
\gamma _{m}$, one has
\begin{eqnarray}
\left\langle \delta q^{2}\right\rangle  &\simeq &\frac{\frac{g_{2}^{2}}{%
4\eta \zeta }+\left( \bar{n}+\frac{1}{2}+\frac{\zeta }{4}\right) \left( 1+%
\frac{\omega _{m}^{2}}{\omega _{fb}^{2}}\right) }{1+g_{2}+\frac{\omega
_{fb}^{2}}{\omega _{m}^{2}}}  \label{qusqcd2} \\
\left\langle \delta p^{2}\right\rangle  &\simeq &\left[ 1+g_{2}+\frac{\omega
_{m}^{2}}{\omega _{fb}^{2}}\right] ^{-1}\left[ \frac{g_{2}^{2}}{4\eta \zeta }%
\left( 1+\frac{g_{2}\gamma _{m}\omega _{fb}}{\omega _{m}^{2}}\right) \right.
\notag \\
&+&\left. \left( \bar{n}+\frac{1}{2}+\frac{\zeta }{4}\right) \left( 1+\frac{%
\omega _{m}^{2}}{\omega _{fb}^{2}}+\frac{g_{2}\gamma _{m}}{\omega _{fb}}%
\right) \right] .  \label{pisqcd2}
\end{eqnarray}%
These results generalize those of Refs.~\cite {quiescence02,courty}, which were restricted to the adiabatic limit $\kappa \to \infty$, and
partially modify their conclusion that one can always achieve ground-state cooling in the large feedback-gain limit $g_{2}\simeq \zeta
\rightarrow \infty $ ($\eta =1$). This is not true in general. From Eq.~(\ref{qusqcd2}) it is clear that, when $g_{2}\simeq \zeta \gg 1,\omega
_{fb}^{2}/\omega _{m}^{2}$, one obtains $\left\langle \delta q^{2}\right\rangle \simeq 1/2+\omega _{m}^{2}/4\omega _{fb}^{2}+(\bar{n}/g_{2})(
1+\omega _{m}^{2}/\omega _{fb}^{2}) $, which implies that the ground state is approached when $\omega _{m}/\omega _{fb} \ll 1$ and
$\bar{n}/g_{2}\ll 1$. However, in the same limit, Eq.~(\ref{pisqcd2}) yields $\left\langle \delta p^{2}\right\rangle \simeq 1/2+( \omega
_{fb}^{2}/4\omega _{m}^{2}) (g_{2}\gamma _{m}/\omega _{fb}) +\bar{n} /g_{2}( 1+\omega _{m}^{2}/\omega _{fb}^{2}+g_{2}\gamma _{m}/\omega _{fb})
$, showing that the feedback bandwidth $\omega_{fb}$ cannot be too large, because otherwise $\left\langle \delta p^{2}\right\rangle$ becomes too
large. The best cooling regime is instead achieved for $\omega _{fb} \sim 3\omega _{m}$ and $g_2 \simeq \xi$ (i.e. $g_{cd}\simeq 2G/\omega
_{m}$), i.e. for \emph{large but finite} feedback gain. This is consistent with the fact that stability imposes an upper bound to the feedback
gain when $\kappa $ and $\omega _{fb}$ are finite. The optimal cooling regime for cold damping is illustrated in Fig.~\ref{Fig4}, where
$n_{eff}$ is plotted versus the feedback gain $g_{cd}$ and the input power $P$, at fixed $ \kappa =3\omega_m$ (bad-cavity condition) and $\omega
_{fb}=3.5\omega_m$. We find a minimum value $n_{eff}\simeq 0.2$, corresponding to an effective temperature $ T\simeq 0.27$ mK for $g_2 \simeq
\xi \simeq 10^4$.  Lower values of $n_{eff}$ can be obtained only if the quality factor $\omega_m/\gamma_m$ is further increased.

\section{Conclusions}

We have developed a general quantum Langevin treatment of radiation-pressure ground-state cooling of a micromechanical oscillator, extending
previous treatments~\cite{brag,Mancini98,courty,quiescence02,marquardt,wilsonrae} to the full parameter range of a stable cavity. Both cavity
self cooling and cold damping are able to approach the ground state, and the comparison of the optimal cooling conditions for both schemes shows
that self cooling is preferable for a good cavity ($\kappa < \omega_m$), while cold damping is more convenient for a bad cavity ($\kappa
> \omega_m$).

\section{Acknowledgements}

This work was supported by the European Commission (programs QAP and SECOQC), by the Italian Ministry for University and Research (PRIN-2005
2005024254), by the Foundational Questions Institute (FQXi grant RFP1-14-06), by the Austrian Science Fund (FWF project P19539), and by the City
of Vienna.

\end{document}